\documentclass[smallextended]{svjour3}
\usepackage[utf8]{inputenc} 

\usepackage{graphicx}

\begin{document}



\title{A toy model for approaching volcanic plumbing systems as complex systems }

\author{Rémy Cazabet \and
        Catherine Annen    \and        
        Jean-François Moyen \and
        Roberto Weinberg \and
}
\institute{R. Cazabet \at
              Univ Lyon, UCBL, CNRS, INSA Lyon, LIRIS, UMR5205, F-69622 Villeurbanne, France
           \and
           C. Annen \at
              Institute of Geophysics of the Czech Academy of Sciences, Prague, Czech Republic          
            \and
           J.F. Moyen \at
              Université Jean-Monnet, Laboratoire Magmas et Volcans, UCA-CNRS-IRD, F-63170 Aubière
               \and
           R. Weinberg \at
              School of Earth, Atmosphere and Environment, Monash University, Clayton, 3800 Victoria, Australia
}

\date{}

\maketitle
Magmas form at depth, move upwards and evolve chemically through a combination of processes. Magmatic processes are investigated by means of field work combined with geophysics, geochemistry, analogue and numerical models, and many other approaches (e.g., \cite{schmeling2019modelling,moyen2021crustal,sparks2019formation}). Volcanism is studied through a combination of monitoring active volcanoes, geophysical imaging, time series of eruptions, as well as investigating eruption products. However, scientists in the field still struggle to understand how the variety of magmatic products arises and there is not consensus yet on models of volcanic plumbing systems.
This is because eruptions result from the integration of multiple processes beneath the eruption centre, rooted in the magma source either in the mantle or lower crust that feeds a complex network of magma bodies linking source and volcano.

Power-law relationships describe eruption magnitudes, frequencies, and durations \cite{deligne2010recurrence}.
Such relationships are typical of non-linear and self-organised systems with critical points. A growing body of evidence indicates that several magma chambers connect before and during eruptions \cite{gualda2022complex}.
This raises new questions about the role of the magmatic network in triggering eruptions and controlling their magnitude and duration. For example, how is the magnitude of an eruption controlled by magma chamber size and network connectivity? How do such networks focus magma migration to one place? Are volcanic eruptions the result of cascading events where perturbation in one magma chamber propagates across the network? 

To answer these questions, it is necessary to develop a dynamic model of the volcanic plumbing system as an interconnected network, and discover the laws that rule the behaviour of both the nodes of the networks and their connections.

\emph{\textbf{Model definition}}
In this work, we investigate the potential of the network approach through a prototype of magma pool interaction and magma transfer across the crust. In network terms, it describes a diffusion process on a dynamic spatial network, in which diffusion and network evolution are intertwined: the diffusion affects the network structure, and reciprocally. The diffusion process and network evolution mechanisms come from rules of behaviour derived from rock mechanics and melting processes. Nodes represent magma pools and edges physical connections between them, e.g., dykes or veinlets. 

\emph{\textbf{Diffusion process}}
Nodes are described by an individual constant $c$, their capacity; variable $v$, the volume of liquid magma inside them; and their pressure $p=f(v,c)$. Edges are described by one constant global parameter $w$, the volume of magma they can transfer per time unit. Top nodes represent volcanoes, outputs of the system. They have infinite $c$. The diffusion process is defined as: a) magma is introduced in the system to one node at the bottom, one unit per simulation step, i.e., $v=v+1$, b) as an effect of time, the volume of magma in the pools decreases, due to solidification, proportionally to their volume, $v$, c) when two pools are connected, magma flows between them until their pressure equilibrates. 

\emph{\textbf{Network evolution process}}
The network evolution is defined as follows: a) when a critical pressure is reached in one pool, an edge links it to a neighbouring pool, with a preference for those located above (distance on the upward vertical axis is reduced by a fact $\alpha$, parameter), to account for the role of melt buoyancy, b) when pressure equilibrates between pools, no magma flows, thus the connection solidifies, i.e. the edge is removed. As more magma enters the system from below, pressure in the system increases and more connections are created, and magma migrates upwards, until it exits the system through top nodes (e.g. volcano). The initial distribution of pools (spatial distribution, maximum volume capacity), $\alpha$, the cooling behaviour, the time allowed for magma migration, are the variables that control system behaviour. 

\begin{figure}[!t]
\centering
\includegraphics[width=0.90\linewidth]{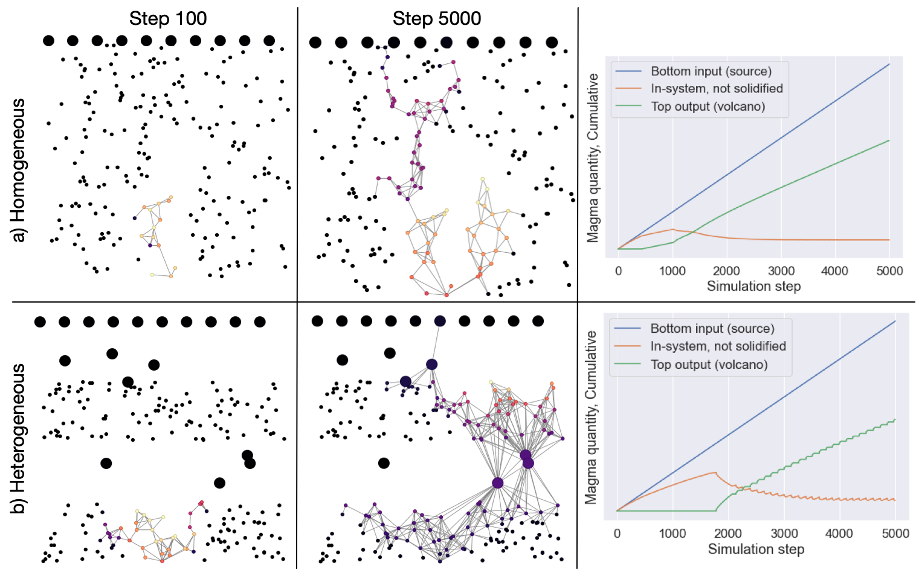}
\caption{Example of the toy model outputs with a) a homogeneous and b) a heterogeneous random distribution of magma pools. Node colours represent pressure, using a relative heat scale (from low to high: black, red, orange, yellow). In b) the four horizontal layers are composed alternatively of either many small capacity pools, or a few large capacity pools. Node size corresponds to the capacity of pools c. Note linear versus pulsating volcano output recorded by the green line (right-hand-side diagrams).}
\label{fig:toy}
\end{figure}

\vspace{0.5cm}
\textbf{Experiments}\\
Two modes of behaviour are shown in Fig. \ref{fig:toy}. All parameters are fixed, except the initial distribution of pools and the maximum volume allowed in each node. Despite a linear magma volume input at the bottom of the system, non-linear behaviours emerge: the homogeneous case (Fig. \ref{fig:toy}a) produces relatively continuous ‘volcanic’ output (green line on right diagram), and the heterogeneous one (Fig. \ref{fig:toy}b) leads to cyclic ‘volcanic’ output. We can track the topology of the network; the activity at each magma pool and connection over time; the average volume of magma as a function of depth, etc.

\vspace{0.5cm}
\textbf{Conclusion}\\
This model is a proof of concept where mechanisms are based on analogies: space and time are not realistic, the system is isothermal, etc. 

Nevertheless, it succeed in showing that a system governed by the same mechanical rules and fed by a linear input of magma at the bottom, can result in nonlinear behaviors at the surface (e.g., episodic volcanic eruptions). 

In future works, we plan to propose a more realistic model, mimicking more closely the natural magmatic plumbing system. We will automatically explore the space of possible node organizations, in order to discover the necessary and sufficient conditions for complex, non-linear behaviors to emerge: are a small number of magma chambers enough? Is the heterogeneity of chambers size more or less important than their physical locations? 

\bibliographystyle{spphys}
\bibliography{volcano}

\end{document}